\documentclass[aps,prb,twocolumn,showpacs,superscriptaddress]{revtex4-1}

\usepackage{graphicx}
\usepackage{amsmath}
\usepackage{amsfonts}
\usepackage{bm}
\usepackage{bbm}
\usepackage{array}

\usepackage{color}
\usepackage{psfrag,pstricks}
\usepackage{multirow}

\begin{document}

\title{Impact of cation-based localized electronic states on the conduction and valence band structure of $\text{Al}_{1-x}\text{In}_x$N alloys}

\author{S. Schulz}
\email{stefan.schulz@tyndall.ie} \affiliation{Photonics Theory
Group, Tyndall National Institute, Dyke Parade, Cork, Ireland}
\author{M.~A. Caro}
\affiliation{Photonics Theory Group, Tyndall National Institute,
Dyke Parade, Cork, Ireland} \affiliation{Department of Physics,
University College Cork, Cork, Ireland}
\author{E.~P. O'Reilly}
\affiliation{Photonics Theory Group, Tyndall National Institute,
Dyke Parade, Cork, Ireland} \affiliation{Department of Physics,
University College Cork, Cork, Ireland}

\date{\today}

\begin{abstract}
We demonstrate that cation-related localized states strongly perturb
the band structure of $\text{Al}_{1-x}\text{In}_x$N leading to a
strong band gap bowing at low In content. Our first-principles
calculations show that In-related localized states are formed both
in the conduction and the valence band in
$\text{Al}_{1-x}\text{In}_x$N for low In composition, $x$, and that
these localized states dominate the evolution of the band structure
with increasing $x$. Therefore, the commonly used assumption of a
single composition-independent bowing parameter breaks down when
describing the evolution both of the conduction and of the valence
band edge in $\text{Al}_{1-x}\text{In}_x$N.
\end{abstract}


\maketitle

The semiconductor alloy $\text{Al}_{1-x}\text{In}_{x}$N has
attracted extensive research interest due to its potential to
improve the performance of electronic and optoelectronic devices.
For high-frequency and high-power microwave applications,
high-electron-mobility transistors based on
$\text{Al}_{1-x}\text{In}_{x}$N/GaN have been considered
recently.~\cite{PaCa2012}
Since the band gaps of AlN and InN are 6.25 eV and 0.69 eV,
respectively,~\cite{Wu2009} low InN compositions are required to
achieve deep UV emission. However, basic and central properties of
$\text{Al}_{1-x}\text{In}_{x}$N alloys are not well known or
understood. For example, when designing optoelectronic devices based
on $\text{Al}_{1-x}\text{In}_{x}$N alloys, knowledge about the
energy gap variation of $\text{Al}_{1-x}\text{In}_{x}$N with
composition $x$ is of central importance. It is common for most
group III-V materials to assume that the band gap of an alloy
$A_{1-x}B_xC$ has a close to linear variation with composition $x$,
with any deviation from linearity described by a small quadratic
term, $-bx(1-x)$, where $b$ is the bowing parameter~\cite{VuMe2001}:
\begin{equation}
E^{ABC}_g(x)=(1-x)E^{AC}_g(x)+xE^{BC}_g(x)-bx(1-x)\,\, .
\label{eq:VCA}
\end{equation}
Here $A$ and $B$ denote cations while $C$ denotes the anion. For
AlInN this simple assumption has been questioned by several groups,
based on experimental data.~\cite{WaMa2008,*SaBe2010,*ScCa2013Apex}
To date, the physical origin of this behavior is unclear.

In this letter we propose and demonstrate that cation-related
localized states strongly perturb the band structure of the III-V
alloy $\text{Al}_{1-x}\text{In}_x$N. We show that In-related
localized states are formed both in the conduction band (CB) and the
valence band (VB) in $\text{Al}_{1-x}\text{In}_x$N for low In
composition $x$, and that these localized states dominate the
evolution of the band structure with increasing $x$. This is
initially surprising -- to date the band structure of all III-V
materials involving alloying between elements from the third to the
fifth row of the periodic table has been described using a single
\emph{composition-independent} bowing parameter $b$.~\cite{VuMe2001}
In addition, the band gap bowing in the two ternary alloys GaInN and
AlGaN is generally well described using
Eq.~(\ref{eq:VCA}),~\cite{VuMe2003,*PeCa2011} but with GaInN showing
some deviations from this simple
description.~\cite{CaSc2013local,MoMi2011} A similar behavior might
therefore also be expected for $\text{Al}_{1-x}\text{In}_x$N. We
note however that AlInN has been reported to have a very large and
composition-dependent bowing parameter, with values of $b$ ranging
from $\sim 2.5$ eV (high In content)~\cite{KiSa97} to 10.3 eV (low
In content).~\cite{AsDa2010} A large bowing parameter has to date
generally been associated with the presence of isoelectronic states
in a semiconductor alloy, such as
$\text{Zn}\text{Te}_{1-x}\text{Se}_{x}$.~\cite{WaSh2000} Using
density functional theory (DFT), we show here that this is also the
case for $\text{Al}_{1-x}\text{In}_x$N, and that the evolution of
the band structure of $\text{Al}_{1-x}\text{In}_x$N for values of
$x$ as large as $\sim 20$\% is dominated by In-related localized
isoelectronic defect levels. This behavior arises due to the marked
difference between the band structure of AlN and InN, including a
difference in energy gap of over 5 eV. Most of this difference in
energy gap between free-standing AlN and InN occurs in the CB, with
the estimated CB offset varying from 4.0 eV~\cite{KiVe2007} to 4.6
eV~\cite{MoMi2011}. This compares with estimated CB offsets of
1.4-2.7 eV between AlN and GaN,~\cite{WaGr96,WeZu96,RiLa99,MoMi2011}
and of 1.6-2.3 eV between GaN and
InN.~\cite{WeZu96,KiVe2008,MoMi2011}

We show here by explicit calculation that the large difference in
the CB energies of InN and AlN leads to the formation of a localized
resonant defect state above the CB edge (CBE) when a single Al atom
is replaced by an In atom in AlN. We further show that the
interaction between In-related localized states and the AlN host
matrix states then leads to a rapid reduction in the CBE energy of
$\text{Al}_{1-x}\text{In}_x$N with increasing In composition $x$,
analogous to the behavior of alloys such as
$\text{ZnTe}_{1-x}\text{Se}_x$.~\cite{WaSh2000}

Turning to the VB states, we note that several studies have shown
how the highest valence states tend to become localized in GaInN
alloys.~\cite{ChUe2006,LeWa2006,WuWa2009,LiLu2010,ChLi2010} We show
that the highest VB states can  be strongly localized in AlInN,
undertaking DFT studies both on ordered and on disordered supercells
(SCs). We first consider ordered SCs, in which each In atom only has
Al atoms as second nearest neighbors. We show that the VB edge (VBE)
energy has a much weaker variation with composition in such
structures compared with that which is observed in structures
containing In pairs and clusters, where an In pair is formed when a
N atom has two In neighbors. We find that even a single In pair in a
SC can strongly shift the VBE energy upwards. It also tends to
localize the highest valence state, consistent with In pairs
introducing localized states below the VBE in
$\text{Al}_{1-x}\text{In}_x$N.

\begin{figure}[t!]
\includegraphics[width=0.8\columnwidth]{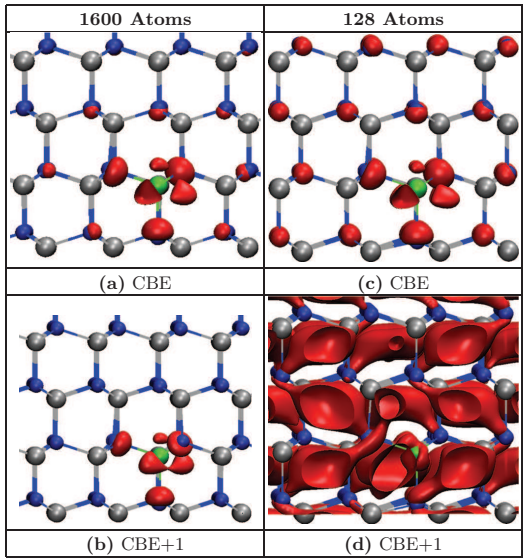}
\caption{(Color online) Charge densities for CBE and CBE+1 at
$\mathbf{k}=\mathbf{0}$ obtained within LDA- (left) and HSE-DFT
(right) calculations. The dark (red) isosurface of the charge
density corresponds to 10\% of the maximum value for the given
eigenstate. SCs contain 1600 (LDA) and 128 (HSE) atoms,
respectively.} \label{fig:CHG_LDA_HSE}
\end{figure}

To analyze the impact of a single, isolated In atom in an AlN
matrix, we first present the results of DFT calculations undertaken
using the local density approximation (LDA) and the plane-wave-based
\emph{ab initio} package \textsc{vasp}.~\cite{KrFu96} SCs with 1600
atoms have been studied. The energy cutoff for the plane waves was
400 eV, the semi-core $d$ electrons of In have been treated as
valence electrons and $\mathbf{k}=\mathbf{0}$ has been chosen. The
structure is relaxed using a valence force field (VFF)
model.~\cite{Mart72,CaSc2013local} Figure~\ref{fig:CHG_LDA_HSE}
shows the calculated charge densities of (a) the CBE and (b) the
first excited CB state (CBE+1) at $\mathbf{k}=\mathbf{0}$.
Figure~\ref{fig:CHG_LDA_HSE}(b) shows explicitly that a single,
isolated In atom introduces a strongly localized state (CBE+1) above
the CBE. Further analysis which we present later of the calculated
participation ratio for each state confirms this behavior. It can
also be seen from Fig.~\ref{fig:CHG_LDA_HSE}(a) that this localized
state hybridizes with the CBE: the wave function (WF) for the lowest
CB state is a linear combination of the AlN CBE state and of the
localized state (CBE+1).

Since LDA calculations underestimate the band gap,~\cite{PeLe83} we
have performed additional DFT calculations within the
Heyd-Scuseria-Ernzerhof (HSE) screened exchange hybrid functional
scheme,~\cite{HeSc2003,*HeSc2004} implemented in \textsc{vasp}. The
same settings as for the LDA calculations have been employed. For
HSE-DFT an exact exchange mixing ratio of $\alpha=0.35$ has been
used. While this setting slightly overestimates the band gap of InN,
$\alpha=0.35$ allows us to describe the energy gap of AlN
accurately.~\cite{MoMi2011} Since we are interested in
$\text{Al}_{1-x}\text{In}_x$N with low In content ($x<0.2$), this
$\alpha$ value should be appropriate. Again the system was relaxed
using the VFF model to address large HSE SCs (64 to 128 atoms). For
the lattice parameters we use the results of the HSE-DFT
relaxation,~\cite{CaSc2012} which have been modified according to
the ratios (relaxed AlN SC vs. relaxed AlInN SC) obtained from the
VFF SC calculations. Here, we treat free-standing structures, with
no strain field arising from an underlying substrate.

The charge densities from HSE-DFT at $\mathbf{k}=\mathbf{0}$ for CBE
and CBE+1 in a 128 atom SC are shown in Fig.~\ref{fig:CHG_LDA_HSE}.
The LDA- (a) and HSE-DFT (c) results show very similar behavior for
the CBE state, both giving a WF which has delocalized CBE character,
mixed with a component which is clearly localized around the In
atom. We have shown previously for dilute nitride alloys that this
mixing of a localized component into the CBE state is a clear
signature for the existence of a localized resonant state above  the
CBE.~\cite{OReLi2009} However, because the localized state is
resonant with the CB, it can be mixed into several different higher
conduction states.~\cite{LiORe2001} Because of this mixing, it is
therefore not possible in the 128-atom SC calculation to associate a
single state above the CBE with the In resonant state: the charge
density of CBE+1 from the HSE-DFT calculation gives a state which is
clearly delocalized. The impact of SC size on localization effects
has been discussed in detail for example in
Ref.~\onlinecite{ChLi2010}; it is common to find that a resonant
localized state cannot be clearly identified for all SC sizes.
However, since HSE- and LDA-DFT give similar results for CBE,
LDA-DFT should also give reliable results for higher CB states. We
conclude that the CBE state is in both cases formed through mixing
of a resonant In defect state with the host matrix CBE state.

\begin{figure}
\includegraphics[width=0.9\columnwidth]{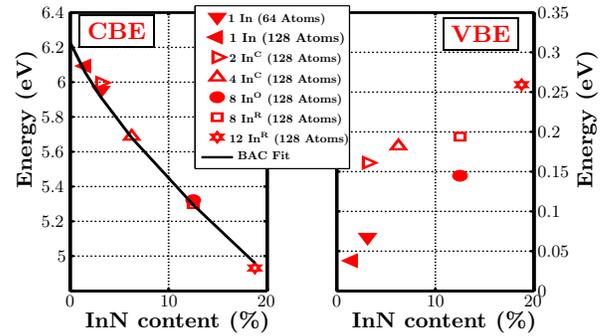}
\caption{(Color online) CBE and VBE as a function of In content $x$
in Al$_{1-x}$In$_x$N supercells. Calculations for isolated In atoms
($\text{I}\text{n}^{\text{O}}$), and for  random
($\text{I}\text{n}^{\text{R}}$) and clustered
($\text{I}\text{n}^{\text{C}}$) configurations. Data is based on
HSE-DFT calculations.} \label{fig:Energy_HSE}
\end{figure}

Having discussed the effect of an isolated In atom on the CB, we now
present HSE-DFT calculations to analyze the evolution of the band
structure with increasing In composition $x$, including the effect
of In pairs. Figure~\ref{fig:Energy_HSE} shows the calculated CBE
and VBE energies as a function of $x$ in
$\text{Al}_{1-x}\text{In}_x$N. We consider structures which include
only isolated In atoms (full symbols; referred to as ordered
structures) as well as structures which include In pairs and
clusters (open symbols). Two types of systems with pairs or clusters
are considered: (i) 128-atom SCs with only two or four In atoms,
where the In atoms share the same N-atom [2 $\text{In}^\text{C}$
(128 Atoms), 4 $\text{In}^\text{C}$ (128 Atoms)], referred to as
clustered structures, and (ii) 128-atom SCs where eight and 12 In
atoms are distributed randomly in the SC [8 $\text{In}^\text{R}$
(128 Atoms), 12 $\text{In}^\text{R}$ (128 Atoms)]. For an analysis
of higher In concentrations, we refer to our recent semi-empirical
tight-binding analysis,~\cite{ScCa2013Apex} which is consistent with
the here presented DFT data.

The calculated variation of the CBE with $x$ can be well described
by a single smooth curve, with $< 50$ meV difference between any two
calculations at the same composition. Turning to the VB, we find
that the VBE energies are consistently higher for structures with In
pairs or clusters (open symbols) compared to the structures with
isolated In atoms only: the VBE shifts upwards by almost $100$ meV
at $x=3.125\%$ between the 64-atom SC with isolated In atoms and the
128-atom SC with an In pair. It is clear that pairs significantly
impact the electronic structure of AlInN.

\begin{table*}
\caption{Normalized CBE, CBE+1 and VBE WF participation ratio
$\overline{\text{PR}}$ for one In atom and an In atom pair in AlN.
LDA-DFT calculations have been performed on a 1600-atom SC; HSE-DFT
are obtained from a SC with 128 atoms.}
\begin{tabular}{c>{\centering\arraybackslash}p{1.75cm}>{\centering\arraybackslash}p{1.75cm}>{\centering\arraybackslash}p{1.75cm}>{\centering\arraybackslash}p{1.75cm}>{\centering\arraybackslash}p{1.75cm}}
\hline
\multirow{2}{*}{$\overline{\textbf{PR}}$} & \multicolumn{3}{>{\centering\arraybackslash}p{5.25cm}}{\textbf{\emph{\hspace*{0.4cm}Isolated In atom}}} & \multicolumn{2}{>{\centering\arraybackslash}p{3.5cm}}{\textbf{\hspace*{0.2cm}\emph{Pair of In atoms}}}\\
\cline{2-6} & \textbf{CBE} & \textbf{CBE+1} & \textbf{VBE} &
 \textbf{VBE}  & \textbf{CBE} \\\hline
\textbf{HSE-DFT}  & \textbf{1.26} & \textbf{1.15} & \textbf{1.23} & \textbf{1.57} & \textbf{1.63}\\
\textbf{LDA-DFT}  & \textbf{1.18} & \textbf{27.56} & \textbf{1.05} &
\textbf{2.51} & \textbf{2.75}\\\hline
\end{tabular}
\label{tab:PRn}
\end{table*}

\begin{figure}
\includegraphics[width=0.9\columnwidth]{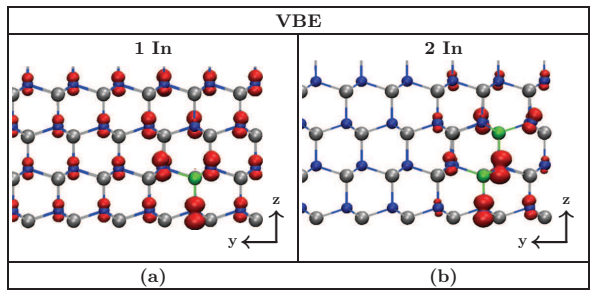}
\caption{(Color online) LDA-DFT (1600-atom SC) results for (a) an
isolated In atom and (b) an In pair. Dark (red) isosurfaces
correspond to 10\% of maximum charge density value.}
\label{fig:VBE_HSE}
\end{figure}

To further analyze the effect of an In pair on the valence band
structure, Fig.~\ref{fig:VBE_HSE} shows the LDA-DFT VBE charge
density calculated in a 1600-atom SC with one In atom (a) and an In
atom pair (b). In AlN, due to the negative crystal field splitting
and neglecting the weak spin-orbit coupling, the VBE at
$\mathbf{k}=\mathbf{0}$ belongs to the irreducible representation
$\Gamma_1$.~\cite{KoDi69} The corresponding VBE Bloch state is
$p_z$-like.~\cite{KoDi69} The calculated VBE charge density in
Fig.~\ref{fig:VBE_HSE} reflects this $p_z$-like character. We find
that even in the one In case, the WF shows signs of localization on
the N sites surrounding the In atom. This localization effect is
enhanced strongly in the two In atom case [cf.
Fig.~\ref{fig:VBE_HSE}(b)]. HSE-DFT calculations for a 128-atom SC
confirm this result (not shown).

To measure the localization effects more quantitatively, we have
calculated the participation ratio (PR), \mbox{$\text{PR}=V\int
|\psi(\mathbf{r})|^4\,\text{d}^3r$}.~\cite{LeWa2006} Here, $V$ is
the volume over which the WF $\psi$ is normalized. The larger the
value of PR for a given volume $V$, the stronger the localization
(for a highly localized state, PR also increases with $V$). Within
LDA- and HSE-DFT we have calculated PR for the CBE, CBE+1 and VBE
states for an isolated In atom and a pair of In atoms sharing a N
neighbor. Table~\ref{tab:PRn} summarizes the normalized
$\overline{\text{PR}}=\text{PR}/\text{PR}^\text{AlN}$ values, where
$\text{PR}^\text{AlN}$ is the PR of the equivalent pure AlN state.
We see from Table~\ref{tab:PRn} that, as soon as one In atom is
introduced in AlN, both HSE- and LDA-DFT results show an increase in
$\overline{\text{PR}}$ for CBE, CBE+1 and VBE. Furthermore, from our
1600-atom SC LDA calculation we find that for CBE+1 $\text{PR}$ is
almost a factor of 30 larger than $\text{PR}^\text{AlN}$. This
confirms the formation of a strongly localized In-related state
above the CBE, as shown in Fig.~\ref{fig:CHG_LDA_HSE}. Likewise, the
introduction of an In pair leads to a strong localization of both
the VBE and the CBE WF, as revealed by the $\overline{\text{PR}}$
values in Table~\ref{tab:PRn}.

To shed further light on these localization effects we focus firstly
on the VBE. The localization of the highest valence state due to an
In pair is related to the orbital character of the highest valence
states (see Fig.~\ref{fig:VBE_HSE}). As discussed above, the VBE
state in AlN is $p_z$-like, localized predominantly on the N atoms.
When an In pair is formed, as in Fig.~\ref{fig:VBE_HSE}(b), the
$p_z$ orbital on the shared N atom interacts with two neighboring In
atoms, leading to a reduced overall bonding interaction for the
given $p_z$ orbital compared to that for a $p_z$ orbital on a N atom
with four Al neighbors. In addition, the interaction between
individual $p_z$ orbitals is stronger along the $z$ direction than
it is in the $x$-$y$ plane. This strong directional dependence
(in-plane versus out-of-plane) of the interactions between $p_z$
orbitals reduces the effective dimensionality of the highest valence
states. This anisotropy, coupled with the reduced bonding
interaction around In sites, supports the formation of localized
states. The impact of the anisotropic interaction can be seen in
Fig.~\ref{fig:VBE_HSE}(b), where there is a rapid reduction of the
VBE probability density in the $x$-$y$ plane in the presence of an
In pair, with a slower decay of the VBE probability density along
the $z$ direction, due to the stronger interaction of $p_z$ orbitals
along that axis.

Recently, valence state localization due to In-N-In chains has been
observed experimentally by Chichibu \emph{et al.} in GaInN
structures.~\cite{ChUe2006} First-principles calculations carried
out for \emph{zinc blende} GaInN on 8-atom~\cite{LeWa2006} and
64-atom SCs with a few selected In configurations~\cite{WuWa2009}
show hole localization when the In impurities are clustered to form
a zig-zag In-N-In-N-In chain~\cite{WuWa2009}. Liu~\emph{et
al.}~\cite{LiLu2010} studied \emph{wurtzite (WZ)} InGaN systems
based on DFT calculations. Special attention was paid to the impact
of uniformly distributed In, an In pair and a condensate where In
atoms congregate together. The authors conclude that short In-N-In
chains lead to a weak localization of valence band
WFs.~\cite{LiLu2010} Our calculations show a similar but even
stronger effect in AlInN, with localization observed even when just
a single In pair is introduced into a large AlN SC. Therefore, we
suggest an experiment similar to that performed by Chichibu \emph{et
al.}~\cite{ChUe2006} for GaInN, to study the localization effects in
AlInN.

The variation of the CBE energy with composition $x$ in III-V alloys
where one of the alloy components introduces resonant defect states
above the CBE can be described by a band-anticrossing (BAC)
model.~\cite{ShWa99,LiORe2004,OReLi2009} Here, the evolution of the
CBE energy is given by the lower eigenvalue of the $2\times2$
Hamiltonian matrix~\cite{ShWa99,LiORe2004}
\begin{equation*}
H(x)=
\begin{pmatrix}
E_\text{D}(x) & V_\text{D,c}\\
V_\text{D,c} & E_\text{c}(x)
\end{pmatrix}
\,\, , \label{eq:BAC_Ham}
\end{equation*}
where $E_\text{D}$ is the resonant defect state energy, which
corresponds to CBE+1 in Fig.~\ref{fig:CHG_LDA_HSE} (b).
$E_\text{c}(x)$ describes the virtual crystal approximation (VCA)
variation of the host matrix CBE state with composition $x$, and
$V_\text{D,c}$ describes the interaction between the levels, which
varies with $x$ as $\beta \sqrt{x}$, where the $\sqrt{x}$ dependence
arises because the interaction is between localized defect and
extended conduction states.~\cite{LiORe99} Thus, the energies of the
CBE ($E_{-}$) and the higher lying resonant state ($E_+$) change
according to
{$E_{\pm}=(1/2)\left(E_\text{c}+E_\text{D}\right)\pm\sqrt{(1/4)\left(E_\text{c}-E_\text{D}\right)^2+V^2_\text{D,c}}$.}
Our 1600-atom LDA-DFT calculations on structures containing a N atom
with 2, 3 or 4 In neighbors show that such pairs and clusters also
give rise to a resonant defect state, which shifts down in energy
but is calculated to still remain above the CBE with increasing
cluster size. We therefore take $E_\text{D}$ to vary as
$E_\text{In}^{0}-\alpha x$, to reflect the increasing number of In
pairs and clusters expected as $x$ increases. Likewise the CBE
varies in the VCA as $E^0_\text{c}-\gamma x$. The solid black line
in Fig.~\ref{fig:Energy_HSE} shows a BAC fit to the variation of the
CBE with composition $x$, calculated with the In resonant state
placed 0.7 eV above the AlN CBE (estimated from the LDA data), and
assuming $\beta=2.5$ eV, and $\alpha = \gamma= 2.3$ eV. Clearly,
with four free parameters, there is a wide range of other values
that could have been chosen. The values used here are comparable
with those previously used for GaN$_x$As$_{1-x}$,~\cite{OReLi2009}
supporting that the evolution of the CBE in
$\text{Al}_{1-x}\text{In}_x$N  is well described using a BAC model
to take account of the interaction between the localized In states
and the host matrix CBE.

Our finding of a BAC interaction in AlInN is further supported by
recent experimental data and DFT calculations showing a marked
reduction in the band gap deformation potential in AlInN for low In
compositions.~\cite{GoKa2010} Localized states such as the In
isoelectronic defect level in AlN typically have a smaller
hydrostatic deformation potential than the CBE. Because the lowest
CB state is formed by mixing such localized state character with the
host matrix CBE, this then leads to a measurable reduction in the
direct gap deformation potential.~\cite{ShWa99}

In conclusion, based on DFT studies we have demonstrated that
cation-induced localized states strongly perturb the band structure
of the III-V alloy $\text{Al}_{1-x}\text{In}_x$N, leading to the
breakdown of the assumption of a single composition-independent band
gap bowing parameter. We have shown that In-related localized states
are formed both in the CB and the VB in
$\text{Al}_{1-x}\text{In}_x$N for low In composition, and that these
localized states dominate the evolution of the band structure with
increasing $x$.

This work was supported by Science Foundation Ireland (project No.
10/IN.1/I2994) and the EU 7th Framework Programme (ALIGHT;
FP7-280587). We thank Peter Parbrook and Emanuele Pelucchi for very
useful discussions.

\end{document}